\documentclass[journal,12pt,onecolumn,draftclsnofoot]{IEEEtran}

\usepackage{amsmath,epsfig,stmaryrd}
\usepackage{color}
\usepackage{dsfont}
\usepackage{caption}
\usepackage{lipsum}
\usepackage{amssymb}
\usepackage{amsthm}
\usepackage{soul}
\usepackage{cite}
\usepackage{setspace}
\usepackage{mathtools}
\usepackage{booktabs}
\usepackage{adjustbox}

\DeclarePairedDelimiter\Floor{\lfloor}{\rfloor}
\newcommand{\erf}{\mathrm{erf}\!}
\renewcommand{\Pr}{\mathbb{P}} 
\usepackage{bm,nicefrac}

\DeclareMathOperator*{\sign}{\mbox{sgn}}

\begin{document}

\title{ \footnote{This work was supported by grant ANR-17-CE40-0020 of the French National Research Agency (ANR).} Energy Optimization of Faulty Quantized Min-Sum LDPC Decoders }

\author{Mohamed Yaoumi$^{1}$, Jeremy Nadal$^{2}$, Elsa Dupraz$^{1}$, Frederic Guilloud$^{1}$, and Fran\c{c}ois Leduc-Primeau$^{2}$ \\  
\small $^1$ IMT Atlantique, Lab-STICC, UMR CNRS 6285, F-29238, France. \\ 
\small $^2$ Department of Electrical Engineering, Polytechnique Montreal, QC, Canada.\\
Emails: $^1$\{mohamed.yaoumi, elsa.dupraz, frederic.guilloud\}@imt-atlantique.fr, $^2$\{jeremy.nadal, francois.leduc-primeau\}@polymtl.ca}
\maketitle
\thispagestyle{empty}

\IEEEpeerreviewmaketitle

\begin{abstract}
The objective of this paper is to minimize the energy consumption of a quantized Min-Sum LDPC decoder, by considering aggressive voltage downscaling of the decoder circuit. Since low power supply may introduce faults in the memories used by the decoder architecture, this paper proposes to optimize the energy consumption of the faulty Min-Sum decoder while satisfying a given performance criterion. The proposed optimization method relies on a coordinate descent algorithm that optimizes code and decoder parameters which have a strong influence on the decoder energy consumption: codeword length, number of quantization bits, and failure probability of the memories. Optimal parameter values are provided for several codes defined by their protographs, and significant energy gains are observed compared to non-optimized setups. 
\end{abstract}


\section{Introduction}

Energy consumption is an important criterion in the design of electronic circuits, and can be greatly reduced by aggressive voltage scaling of the circuit. Low power supply may however introduce faults in the computation operations and memories of the circuit~\cite{gupta2012underdesigned}. 
In this paper, we address this issue in the area of channel coding, and more specifically for Low Density Parity Check (LDPC) codes. The objective is to find the best compromise between decoder circuit energy consumption and LDPC decoding performance under circuit faults.

Two energy consumption models are provided in~\cite{ganesan2015towards} for non-faulty LDPC decoders: the first model estimates the decoding complexity, while the second evaluates the wire length in the circuit.
Then,~\cite{nguyen2016non} introduces a method to minimize the alphabet size of quantized messages exchanged in the decoder, aiming to lower the memories energy consumption.  
Finally,~\cite{smith2010design}, proposes to optimize the code rate and irregular code degree distribution in order to minimize the decoder complexity and therefore its energy consumption.

In addition, the performance of LDPC decoders implemented on faulty hardware was widely studied in the literature. In~\cite{huang2013gallager} the authors assume that the LDPC decoder is subject to both transient and permanent errors. Transient errors make faulty gates or memory units provide an erroneous output from time to time with a non-zero probability. Permanent errors make a fraction of the gates and memories stuck at the same output. When dealing with energy consumption issues, we consider process diversity strategies, where the permanent errors turn into transient error~\cite{leduc2018modeling}.  
The authors in~\cite{varshney2011performance,balatsoukas2014density,ngassa2015density,leduc2018modeling} theoretically investigate the effect of transient errors on various LDPC decoders, such as Gallager A and B or quantized Min-Sum. 
However, none of these works relate the amount of faults introduced in the decoder to its energy consumption.

In this work, our objective is to minimize the energy consumption of a faulty LDPC decoder while satisfying a given performance criterion. For this, we consider protograph-based LDPC codes and quantized Min-Sum decoders for their easy hardware implementation~\cite{karkooti2004semi}.
In~\cite{marchand2011architecture}, it is shown that memories represent the largest part in term of area and energy consumption of circuits of LDPC decoders.
Therefore, in this paper, we assume that the decoder faults are introduced in the memory units, and we consider the noise model introduced in \cite{chatterjee2016energy}, that relates the noise in the stored bits with the energy consumption of a memory cell. 
 
 To estimate the LDPC decoder energy consumption, we update the non-faulty memory energy model of \cite{yaoumi2019optimization}, in order to apply it to faulty decoders. This energy model depends on several code and decoder parameters, such as the protograph, the noise level, the number of quantization bits for the messages, the codeword length, and the number of iterations performed by the decoder. In order to properly evaluate the proposed energy model, we consider the method of~\cite{yaoumi2020energy} which relies on Density Evolution (DE) in order to estimate the average number of decoder iterations required for a given codeword length. 
 Then, protograph optimization being a difficult problem itself~\cite{yaoumi2019optimization}, we consider a fixed prtograph, and we propose a method to optimize both the codeword length and the decoder parameters (number of quantization bits and noise level) in order to minimize the decoder energy consumption. This method is based on a coordinate-descent algorithm that successively optimizes each parameter, assuming that the other ones are fixed, and repeats the process over several iterations.  Simulation results provide the values of optimized parameters for several protographs, and show the energy gains compared to non-optimized decoders. 
 

\section{LDPC codes and decoders} \label{sec:LDPC}
We consider a codeword $\mathbf{x}$ of length $N$ to be transmitted over an additive white Gaussian noise (AWGN) channel of variance $\sigma^2$, with binary phase-shift keying (BPSK) modulation. We use $y_i$ to denote the $i$-th channel output, and $x_i \in \{-1,1\}$ to denote the $i-$th modulated coded bit. 
The channel Signal-To-Noise Ratio (SNR) is defined as $\xi = 1/\sigma^2$.
In this section, we introduce our notation for protograph-based LDPC codes, and describe the considered faulty quantized Min-Sum decoder.

\subsection{Protograph-based LDPC codes}
LDPC codes are represented by a sparse $M\times N$ parity-check matrix $H$
where $h_{j,i}$ is the coefficient in the $j$-th row  and $i$-th column of $H$. 
Assuming that ${H}$ is full rank, the code rate is $R=K/N$, where $K=N-M$ is the information length.
The matrix $H $ can also be represented by a Tanner graph with N variable nodes and M check nodes. The i-th Variable Node (VN) $v_i$ and the  j-th Check Node (CN) $c_j$ are connected in the Tanner graph if $h_{j,i}=1$. 
We use $\mathcal{N}_{c_j}$  to denote the set of all VNs connected to CN $c_j$, and we use $\mathcal{N}_{v_i}$ to denote the set of all CNs connected to VN $v_i$. 


In this work, we consider LDPC codes constructed from protographs~\cite{fang2015survey}. A protograph $S$  is a matrix of size $m \times n$  that gives the number of connections between each VN and CN in the reduced Tanner graph representing the protograph. We can construct an LDPC code of length $N$ by first copying the protograph $Z$ times where $Z=N/n$ is called the lifting factor, and then by interleaving the edges to get the parity check matrix $H$.   


\subsection{Faulty quantized Min-Sum decoder}
\label{sec:faultyDecoder}
In this paper, we consider a quantized offset Min-Sum  decoder~\cite{balatsoukas2014density,ngassa2015density,leduc2018modeling}, implemented with the architecture proposed in~\cite{dupraz2018low}. For simplicity, no pipeline stages are considered, which corresponds to a row-layered scheduling. This enables to use fewer decoding iterations and reduces the size of the circuit. 
The decoder messages are quantized on $q$ bits and between values $-Q$ and $Q$, where $Q=2^{q-1}-1$. We consider the following quantization function:
\begin{equation}
\label{eq:quantization}
\Delta(x)=\sign(x)\min\left(Q, \Floor*{ |x|+\frac{1}{2}}\right) \,, 
\end{equation} 
where $\sign(x)=1$ if $x\geq 0$, and $\sign(x)=-1$ if $x<0$.

Since memory units are responsible for a large part of the decoder energy consumption~\cite{marchand2011architecture}, we assume that faults are introduced during memory read operations in the faulty decoder. The error model corresponds to XOR-wise operation $\oplus$ between the memory read port output and a noise term represented by independent and identically distributed random variables $\mathcal{B}$. These random variables can equivalently be represented on $q$~bits as $(b_1,\cdots,b_q)$. We assume that the random variables $b_k$ are independent and identically distributed, and that each $b_k$ follows a Bernoulli distribution with parameter $\epsilon$. 

In order to initialize the decoder, we compute log-likelihood ratio (LLR) $r_i$ for each received value $y_i$ as:
 \begin{equation}
r_i = \alpha\log\left(\frac{\Pr\left(x_i=1 | y_i \right)}{\Pr\left(x_i=-1 | y_i \right)}\right) = \frac{2\alpha y_i}{\sigma^2},
\end{equation}
where  $\alpha$ is a scaling parameter.
In the  architecture proposed in~\cite{dupraz2018low}, the
VN messages $\beta^{(\ell)}_{i}$ at iteration $\ell \in \llbracket 1,L \rrbracket$ are updated as
\begin{equation} \label{eq:variable_update}
\beta^{(\ell)}_{i} \leftarrow \beta^{(\ell)}_{i} + \sum_{j\in \mathcal{N}_{v_i}}{\left(\gamma^{(\ell)}_{j\rightarrow i } - \left(\gamma^{(\ell-1)}_{j\rightarrow i } \oplus \mathcal{B}_{j}^{(\ell-1)} \right) \right)}, 
\end{equation}
with $\beta^{(0)}_{i} = \Delta(r_i)$,  $\gamma^{(\ell)}_{j\rightarrow i}$ represents the message sent from the CN $c_j$ to the VN  $v_i$ at iteration $\ell$, and $\mathcal{B}_{j}^{(\ell-1)}$ is the noise introduced when the messages $\gamma^{(\ell-1)}_{j\rightarrow i}$ are read from their dedicated memory. The messages $\beta^{(\ell)}_{i}$ are quantified on $q + q_s$ bits, with $q_s=\lceil\log_2( \max d_{v_i} +1)\rceil$, in order to avoid any saturation issue when writing the message $\beta^{(\ell)}_{i}$ into the memory. In the considered architecture, the message $\beta^{(\ell)}_{i\rightarrow j}$ sent from the VN $i$ to the CN $j$, at iteration $\ell$, is calculated during the CN update, which is as follows:
\begin{eqnarray}
\beta^{(\ell)}_{i\rightarrow j} &=&  \left(\beta^{(\ell)}_{i} \oplus \mathcal{B}_{i}^{(\ell)} \right) - \left(\gamma^{(\ell-1)}_{j\rightarrow i } \oplus \mathcal{B}_{j}^{(\ell-1)} \right) \label{eq:msg_update} \\
    \label{eq:check_update} \gamma^{(\ell)}_{j\rightarrow i} &=&\left({ \prod_{i' \in \mathcal{N}_{c_j} \setminus \{i\}}{\sign \left({\beta^{(\ell)}_{i'\rightarrow j}}\right) } }\right) \times \max\left[ \min _{j' \in \mathcal{N}_{c_j} \setminus \{i\}}\left|\beta^{(\ell)}_{i'\rightarrow j}\right|-\lambda,0\right],
\end{eqnarray}
where $\lambda$ is an offset parameter, and where $\mathcal{B}_i^{(\ell)}$ represents the noise introduced when reading the memory where the variable-node messages are stored. 
The decoder stops when a stopping criterion is satisfied, or when the maximum number of iterations $L$ is reached.

\section{Finite-length performance evaluation } \label{sec:Perf}

DE~\cite{richardson2001design,richardson2002multi} allows to estimate the error probability $p_{e_\infty}^{(\ell)}(\xi)$ of an LDPC decoder, for a given protograph $S$ and at given SNR $\xi$ and iteration number $\ell$. However, DE calculates $p_{e_\infty}^{(\ell)}(\xi)$ under the assumption that the codeword length tends to infinity. As an alternative,~\cite{leduc2016finite} provides a method to estimate the  error probability $p_{eN}^{(\ell)}(\xi)$ of an LDPC decoder at finite length $N$.  This method estimates the error probability $p_{eN}^{(\ell)}(\xi)$ as
\begin{equation}\label{eq:pe_average}
  p_{e{_N}}^{(\ell)}(\xi)=\int^{\frac{1}{2}}_0 p_{e{_\infty}}^{(\ell)}\left( x \right) \phi_\mathcal{N}\!\left(x;\, p_0,\frac{p_0(1-p_0)}{N}\right)dx  .
\end{equation}
In this expression  $p_0= \frac{1}{2}-\frac{1}{2}\erf\left(\sqrt{\xi/2}\right)$, and $p_{e_{\infty}}^{(\ell)}\left(x\right)$ is the error probability evaluated with standard DE at SNR value $ 2 (\erf^{-1}(1-2x))^2$. The function $\phi_\mathcal{N}(x;\mu,v^2)$ is the probability density function of a Gaussian random variable with mean $\mu$ and variance $v^2$.

For simplicity, we implemented the DE equations by considering a flooding scheduling. Following~\cite{sharon2007efficient}, we empirically obtained the same error probabilities as a row-layered scheduling, given that the number of iterations is doubled.
In addition, the DE equations were derived by considering that the memory faults are introducing after computation of the  check-to-variable messages $\beta^{(\ell)}_{i\rightarrow j}$, as in~\cite{balatsoukas2014density}. This slightly differs from the hardware decoder of~\cite{dupraz2018low} described in Section~\ref{sec:LDPC}, where the faults are introduced when the VN messages $\beta^{(\ell)}_{i}$  are read. However, we observed through simulations a negligible difference on the obtained error probabilities. 



Then, in order to evaluate the decoder energy consumption, we need to estimate the number of iterations required by the decoder. 
Therefore, we use the method of~\cite{yaoumi2020energy} to evaluate the average number of iterations $L_N(\xi)$ at codeword length $N$:
\begin{equation}\small \label{eq:estimateL_FL}
L_N(\xi) = \int_{0}^{\frac{1}{2}} \left(\sum_{\ell=1}^L  R^{(\ell-1)}_{\infty}(x) \right) \Phi_N\left(x;p_0,\frac{p_0(1-p_0)}{N}\right) dx.
\end{equation}
where $R^{(\ell-1)}_{\infty}(x) = 1 - (1-p_{e_{\infty}}^{(\ell)}\left(x\right))^N$.
As for the error probability $p_{e{_N}}^{(\ell)}(\xi)$ defined in~\eqref{eq:pe_average}, the expression of $L_N(\xi)$ takes into account the channel variability, but does not evaluate the effect of cycles onto the decoder performance.  
However, as shown in \cite{leduc2016finite,yaoumi2019optimization}, these two formula accurately predict the finite-length decoder performance for long codewords.

\section{Energy Model} \label{sec:energy}
This section introduces the memory energy model we consider for the faulty Min-Sum decoder described in Section~\ref{sec:LDPC}. 

\subsection{Faults-vs-energy model}
We consider the generic faults-vs-energy model introduced in~\cite{chatterjee2016energy}, that relates the memory failure parameter $\epsilon$ to the energy level $e_g$ as 
\begin{equation}\label{eq:noiseModel}
    \epsilon= \epsilon_0\exp{(-ce_g)},
\end{equation}
where $\epsilon_0$ and $c$ are positive constants that depend on the circuit technology.
In order to specify this model in our setup, we express the energy $E_\mathrm{bit}$ of writing one bit in memory as
\begin{equation}\label{eq:Ebit}
E_\mathrm{bit} = e_g E_0 ,
\end{equation}
where $e_g$ takes values in $[0,1]$, and $E_0$ is referred to as the nominal energy.
 Therefore, $e_g=0$ means that the device does not consume any energy, and $e_g=1$ means that the device operates with nominal energy $E_0$. 

We now give an example on how to set the parameters $c$, $\epsilon_0$, and $E_0$, depending on the technology.
First, for $e_g=0$ we consider a failure probability of $\frac{1}{2}$, which gives that $\epsilon_0=\frac{1}{2}$. 
Then, for a typical $65$nm SRAM cell, the failure probability at a nominal voltage is $ 10^{-7}$ \cite{dreslinski2010near}, which gives $c=12$.  
The value of $E_0$ can be estimated based on \cite{horowitz:2014}, where $10$ pJ is the storage energy for a $64$-bit access from a $8$Kb cache, which gives $E_0=0.156$ pJ. These values will be considered in our simulations.

\subsection{Memory energy model}
The energy model proposed in~\cite{yaoumi2019optimization} estimates the overall memory energy consumption of a non-faulty quantized Min-Sum decoder by counting the total number of bits written into memory during the decoding process. 
In~\cite{yaoumi2019optimization}, the total number of bits written in memory is evaluated from the facts that: (i) at a VN, the message $\beta_i$ is stored on $q+q_s$ bits,  (ii) since we are using a row-layered scheduling, a VN updates its messages every time one of its neighboring check nodes is updated, (iii) at a CN, $1$ bit is stored for the sign of the output message, and two minimum absolute values of $q-1$ bits each are stored.

Here, we consider the memory energy consumption per information bits, in order to properly capture the effect of increased codeword length $N$.  As a result, the following energy model will be considered in the optimization:
\begin{equation}   \label{eq:memoryEnperK}
  \mathcal{E} =  \frac{E}{K}=\frac{L_N(\xi) }{Rn}E_\mathrm{bit} \left(\sum^n_{i=1}d_{v_i}\left(q+q_s\right) +\left(1-R\right)\left(\sum^m_{j=1}\left(2q+d_{c_j}-2\right)\right)\right).
\end{equation}
where $E$ is the total memory energy consumption of the decoder, $L_N(\xi)$ is the average number of iterations given in~\eqref{eq:estimateL_FL} for codeword length $N$, and
 $E_\mathrm{bit}$ can be expressed with respect to the failure probability $\epsilon$ from~\eqref{eq:noiseModel} and~\eqref{eq:Ebit}.

\section{Energy Optimization} \label{sec:optim}
We now propose an optimization method to minimize the decoder energy consumption while satisfying a certain performance criterion. 

\subsection{Optimization problem}
As a performance criterion for the optimization, we fix a target error probability $p_{e}^{\star}$ to be reached at a target SNR value $\xi^{*}$.
For simplicity, we assume that the code rate $R$ and the protograph $S$ are fixed. We propose to minimize the energy consumption $\mathcal{E}$ with respect to the quantization level $q$, the noise parameter $\epsilon$, and the codeword length $N$, while satisfying the performance criterion.
The optimization problem can then be formulated as 
\begin{equation}  \label{eq:optim_pb}
\min_{ \epsilon,q,N} \mathcal{E}(\xi^{*},q,\epsilon,N)\ \ \mbox{s.t.} \  \  p_{e,\text{opt}}(\xi^{*},\epsilon,q,N) < p_{e}^{\star} 
\end{equation}
where $$p_{e,\text{opt}}(\xi^{*},\epsilon,q) = \min_{\alpha,\lambda} p_{e_N}^{(L)}(\xi^{*},\epsilon,q,N).$$ In the above optimization problem, $\mathcal{E}(\xi^{*},q,\epsilon,N)$ is given by~\eqref{eq:memoryEnperK}
and $p_{e_N}^{(L)}(\xi^{*},\epsilon,q,N)$ is calculated from~\eqref{eq:pe_average}.
In addition, $p_{e,\text{opt}}(\xi^{*},\epsilon,q)$ gives the minimum error probability that can be reached by optimizing the scaling parameter $\alpha$ and the offset parameter $\lambda$. 





\subsection{Optimization method}
The optimization problem  \eqref{eq:optim_pb} is difficult to solve because it involves discrete parameters $q$ and $N$.
In addition, it is computationally expensive to evaluate the number of iterations $L_N$ and the error probability $p_{e,N}^{(\ell)}$ for given parameters $q,\epsilon,N$, because this requires to numerically evaluate integrals.  
Therefore, we want to lower the number of evaluations of these terms. 

In order to solve the optimization problem~\eqref{eq:optim_pb}, we first define search intervals for the parameters $q,N,\epsilon$ involved in the optimization. 
First, according to Section~\ref{sec:energy}, the continuous parameter $\epsilon$ lies in the interval $[0,1]$. 
Then, we assume that discrete parameters $q$ and $N$ take values in the sets $\llbracket q_{\min}, q_{\max}\rrbracket$ and $\llbracket N_{\min}, N_{\max}\rrbracket$, respectively.
The range of values for $q$ and $N$ must be selected so as to satisfy the performance criterion at least for the largest values $N_{\max}$ and $q_{\max}$. For instance, in our simulations, we set $q_{\max} = 8$, since for this case, the performance of the quantized Min-Sum decoder is almost the same as the performance of the non-quantized decoder. 

Once the search intervals are set, we then perform a coordinate descent optimization, which consists of optimizing alternatively each of the three parameters $\epsilon$, $q$, and $N$, over several iterations. Since we consider a constrained optimization problem, we verify at each iteration that the selected parameters meet the performance criterion of the optimization problem. For this reason, we first initialize our algorithm with the three parameters $\epsilon^{(0)} = \epsilon_{\max}$, $q^{(0)} = q_{\max}$, and $N^{(0)} = N_{\max}$. Then, at iteration $i \in \llbracket 1,I\rrbracket$, we successively solve the following three optimization problems:
\begin{enumerate}
\item Given $\epsilon^{(i-1)}$ and $ N^{(i-1)}$, solve  \begin{equation}\label{eq:optQ} q^{(i)}  =  \arg\min_{q} \mathcal{E}(\xi^{*},q)\ \ \mbox{s.t.} \  \  p_{e,\text{opt}}(\xi^{*}) < p_{e}^{\star} 
\end{equation}
 \item Given $\epsilon^{(i-1)}$ and $q^{(i)}$, solve
 \begin{equation}\label{eq:optN}
  N^{(i)}  =  \arg\min_{N} \mathcal{E}(\xi^{*},N)\ \ \mbox{s.t.} \  \  p_{e,\text{opt}}(\xi^{*},N) < p_{e}^{\star} 
  \end{equation}
   \item Given $q^{(i)}$ and $N^{(i)}$, solve
   \begin{equation}\label{eq:optEps}
 \epsilon^{(i)}  =  \arg\min_{q} \mathcal{E}(\xi^{*},\epsilon)\ \mbox{s.t.} \  \  p_{e,\text{opt}}(\xi^{*},\epsilon) < p_{e}^{\star} 
\end{equation}
\end{enumerate}
In~\eqref{eq:optQ}, the parameter $q$ is optimized by exhaustive search since the search interval is small. Then, for the optimization of $N$ and $\epsilon$ in~\eqref{eq:optN} and~\eqref{eq:optEps}, we retain the parameters that satisfy the performance criterion $p_{e,\text{opt}}$ and minimize the energy $\mathcal{E}$ among a certain number of values between $N_{\min}$ and $N_{\max}$ and between $\epsilon_{\min}$ and $\epsilon_{\max}$, respectively.  
To further reduce the computation time, we first evaluate the performance criterion $p_{e,\text{opt}}$, and then evaluate the corresponding energy $\mathcal{E}$ only if the performance criterion is satisfied. 

Finally, the coordinate descent approach guarantees that the energy criterion is reduced at each iteration. 
It also ensures that the final solution satisfies the performance criterion. However, there is a risk that the algorithm falls onto a local minimum. This issue is discussed in the next section, in which we evaluate through numerical simulations the proposed optimization method. 

  \section{Numerical  results}\label{sec:simu}
  In this section, we consider four different protographs given in Table~\ref{tab:optab}, all  with parameters $m=2$, $n=4$, and code rate $R=0.5$. The protographs $S_{17}$ and $S_{36}$   were constructed using 
     a genetic algorithm called Differential Evolution~\cite{richardson2001design} that optimizes protographs for performance only. When applying Differential Evolution, the protographs were optimized by considering a large quantization level $q=8$ in order to get a performance very close  to the non-quantized decoder. We also consider the protographs $S_{m}$ and $S_{c}$ that were obtained in \cite{yaoumi2019optimization} by optimizing the decoder energy consumption.  
     
  For the four protographs, we set $p_{e}^{\star} =10^{-3}$ as the target error probability to be achieved at the SNR $\xi^*=1.45$ dB. We then find the optimum parameters $q_{\text{op}}$, $\epsilon_{\text{op}}$ and $N_{\text{op}}$ from the method proposed in Section~\ref{sec:optim}, with $I=3$ iterations.  
  The optimal  parameters are provided in Table~\ref{tab:optab}, along with the corresponding energy value $\mathcal{E}_{\min}$.
  In order to verify over these four protographs that the coordinate descent algorithm did not fall onto a local optimum, we also solved the optimization problem~\eqref{eq:optim_pb} by exhaustive search, and found the same optimum values given in Table~\ref{tab:optab} for the three parameters $q,\epsilon,N$.
  We further observe that, for every protograph, the minimum energy  is achieved for the same quantization level $q_{\text{\text{op}}}=5$ and that the optimal failure   probabilities are close to each other, \emph{i.e.}, $1.02\times10^{-5}<\epsilon_{\text{op}}<3.38\times10^{-5}$, which roughly corresponds to using between 80\% and 85\% of the nominal energy $E_0$. On the other hand, the optimal code length $N_{\text{op}}$ strongly depends on the considered protograph.

We also compare the obtained minimum energy values with respect to two setups in which we optimize only a part of the parameters. In Table~\ref{tab:optab}, $\mathcal{E}_{q_{\text{op}}}$ gives the minimum energy value when only $q$ is optimized, and when $N=10000$ and $e_g =1$. And $\mathcal{E}_{N_{\text{op}},q_{\text{op}}}$ gives the minimum energy value when both $q$ and $N$ are optimized, and when $e_g =1$. We observe an energy gain of $10$ to $15\%$ in the case where all the parameters are optimized.

 \begin{table*}[t]
\small
    \centering
    \caption{Minimum energy value $\mathcal{E}_{\min}$ and optimal parameters for the four considered protographs. The left part of the table represents the case where the three parameters ($q$, $N$, $\epsilon$) are optimized. The right part gives energy values $\mathcal{E}_{q_{\text{op}}}$ and $\mathcal{E}_{N_{\text{op}},q_{\text{op}}}$, when only $q$ is optimized, and when only $q$ and $N$ are optimized, respectively. }

\begin{tabular}{|c|cccc|ccc|cc|}
\hline
Protograph & $\mathcal{E}_{\min}$ & $e_{g_{\text{op}}}$ & $q_{\text{op}}$ &$N_{\text{op}}$ & $\mathcal{E}_{N_{\text{op}},q_{\text{op}}}(e_g=1)$ & $q_{\text{op}}$ & $N_{op}$ & $\mathcal{E}_{q_{\text{op}}}(N=10000, e_g=1)$ & $q_{\text{op}}$ \\
\hline
     
     $S_{17}=\begin{bmatrix} 2& 3& 1& 1 \\0& 1& 4& 1 \end{bmatrix}$ & $79$ pJ& $0.82$& $5$ & $3160$ & $88$ pJ& $5$& $3080$ & $91$ pJ & $5$\\
     $S_{36}=\begin{bmatrix} 2& 1& 2& 3 \\1& 4& 0& 1 \end{bmatrix}$ & $95$ pJ & $0.82$& $5$ & $6320$  & $105$ pJ& $5$&$8020$ & $105$ pJ & $5$\\
     $S_m=\begin{bmatrix}3& 2& 1& 2 \\0& 1& 1& 4\end{bmatrix}$ & $87$ pJ& $0.80$& $5$ & $ 6600$  & $99$ pJ& $5$& $10000$ &  $99$ pJ & $5$\\ 
     $S_c=\begin{bmatrix}0& 1& 2& 5 \\2& 2& 0& 2\end{bmatrix}$ & $91$ pJ & $0.84$ & $5$ & $4160$ & $99$ pJ &$5$& $4080$ & $103$ pJ & $5$\\ 
\hline
\end{tabular}\label{tab:optab}
\end{table*}
\normalsize

 \begin{figure}
	\centering
	\includegraphics[width=10cm]{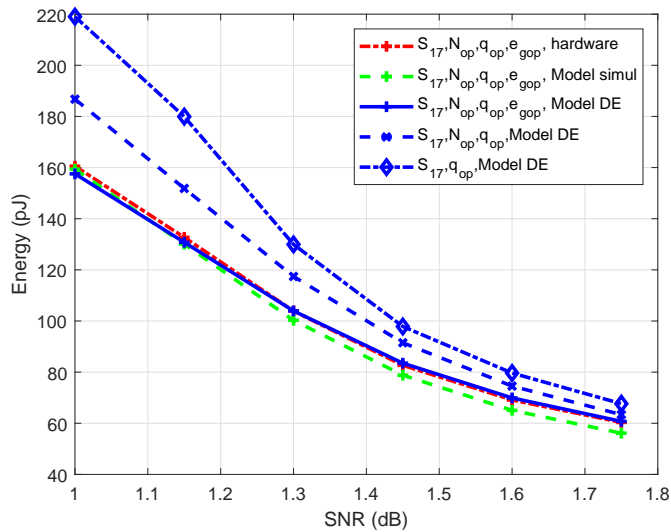}
	\caption{Energy values $\mathcal{E}_{\min}$, $\mathcal{E}_{q_{\text{op}}}$, and  $\mathcal{E}_{N_{\text{op}},q_{\text{op}}}$, with respect to SNR, for the protograph $S_{17}$.}
	\label{fig:EMK}
\end{figure}

\begin{figure}
	\centering
	\includegraphics[width=10cm]{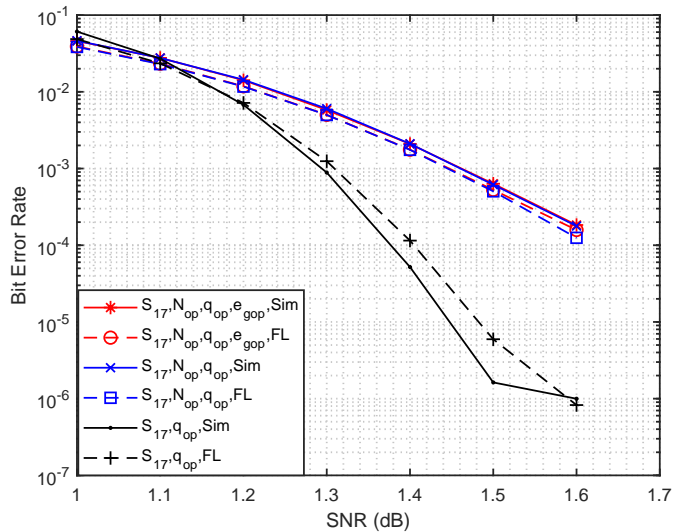}
	\caption{BER with respect to SNR of the protograph $S_{17}$, evaluated from the finite-length DE method and from Monte Carlo simulations.  }
	\label{fig:Pen}
\end{figure}
    We now focus on the protograph $S_{17}$.  Figure~\ref{fig:EMK} gives the values of $\mathcal{E}_{\min}$, $\mathcal{E}_{q_{\text{op}}}$ and  $\mathcal{E}_{N_{\text{op}},q_{\text{op}}}$ with respect to SNR, evaluated from~\eqref{eq:memoryEnperK}. As explained in Section~\ref{sec:faultyDecoder}, the memory fault model considered for DE is simplified, and slightly differs from the hardware implementation. To validate the accuracy of the model, the energy $\mathcal{E}_{\min}$ is evaluated from the average number of iterations $L_N(\xi)$ obtained through
\begin{enumerate}
    \item simulation of the decoder using the hardware memory fault model,
    \item simulation of the decoder using the simplified model, 
    \item DE (equation~\eqref{eq:estimateL_FL}) using the simplified model.
\end{enumerate}
In every considered cases, we use the parameters $q$, $N$, and $\epsilon$ provided in Table~\ref{tab:optab} which were optimized for the target SNR $\xi^{\star}$. We see that although the optimization is performed at one single SNR value, the performance order is preserved at any SNR. Furthermore, the energy obtained with the simplified memory fault model demonstrates a good match when compared to the simulated models.
The figure also confirms the energy gain at optimizing together the three parameters. 

At the end, for protograph $S_{17}$, Figure~\ref{fig:Pen} shows the bit error rate (BER) with respect to SNR, evaluated both from the finite-length method of Section~\ref{sec:Perf} and from Monte-Carlo simulations. Again, the three sets of parameters leading to $\mathcal{E}_{\min}$, $\mathcal{E}_{q_{\text{op}}}$, and  $\mathcal{E}_{N_{\text{op}},q_{\text{op}}}$, are considered. We first observe that the finite-length method of Section~\ref{sec:Perf} accurately predicts the decoder error probabilities, with a maximum gap of $0.1$ dB. In addition, as expected, we see that the case where the three parameters are optimized shows a degraded performance. 
Finally, we conclude that optimizing the decoder parameters allows to reduce the decoder energy consumption, with respect to the decoding performance criterion.

\section{Conclusion}
In this paper, we introduced an energy model for faulty quantized Min-Sum decoders. We then proposed a method to optimize the number of quantization bits, the code length, and the failure probability, in order to minimize the energy consumption while satisfying  a given decoding performance criterion. Simulation results show that using the optimal parameters greatly reduces the energy consumption 
while satisfying the performance criterion.  


\bibliographystyle{IEEEtran}
\bibliography{biblio}

\end{document}